
\font\elvrm=cmr10  scaled\magstep1
\font\elvmi=cmmi10  scaled\magstep1
\font\elvsy=cmsy10  scaled\magstep1
\font\elvex=cmex10  scaled\magstep1
\font\elvit=cmti10  scaled\magstep1
\font\elvsl=cmsl10  scaled\magstep1
\font\elvbf=cmbx10  scaled\magstep1

\font\ninerm=cmr9         \font\eightrm=cmr8          \font\sixrm=cmr6
         \font\eighti=cmmi8          \font\sixi=cmmi6
        \font\eightsy=cmsy8         \font\sixsy=cmsy6

\font\nineit=cmti9        
        
\font\ninebf=cmbx9

\newcount\textno
\newcount\scriptno
\global\textno=10
\global\scriptno=7

\textno=12\scriptno=7
\normalbaselineskip=13dd  
\textfont0=\elvrm  \scriptfont0=\eightrm  \scriptscriptfont0=\sixrm
\textfont1=\elvmi  \scriptfont1=\eighti  \scriptscriptfont1=\sixi
\textfont2=\elvsy  \scriptfont2=\eightsy  \scriptscriptfont2=\sixsy
\textfont3=\elvex  \scriptfont3=\elvex    \scriptscriptfont3=\elvex
\textfont4=\elvit  \textfont5=\elvsl      \textfont6=\elvbf
\def\rm{\fam0\elvrm}                      
\def\sl{\fam5\elvsl}                      \def\bf{\fam6\elvbf}
\setbox\strutbox=\hbox{\vrule height8.5dd depth4dd width0dd}
\normalbaselines\rm
{\vskip-2cm
\headline={\tt 
$\backslash$ 
$\backslash$ 
}\vskip2cm}

\nopagenumbers
\def\rightheadline{{\eightrm\hfil
\hfil}\folio}
\def\leftheadline{\folio\eightrm\hfil
\hfil}
\headline={\vbox to 0pt{\vskip-1.82pc\line{\vbox to 8.5pt{}
\ifodd\pageno\rightheadline \else\leftheadline\fi}\vss}}

\vsize19.2cm \hsize13.4cm \baselineskip=9.6mm
\voffset2.5pc\hoffset2.5pc
\overfullrule=0pt
\centerline{\bf Conceptual analysis of quantum history theory}
\vskip1cm
\centerline{Giuseppe Nistic\`o}
\par
\centerline{\nineit Dipartimento di Matematica,
Universit\`a della Calabria -- Istituto Nazionale di Fisica Nucleare,} 
\par
\centerline{\nineit Gr.c. Cosenza, \quad 87036 Arcavacata, Rende (CS), Italy. gnistico@unical.it}
\vskip6mm\noindent
{\bf A}{\ninebf BSTRACT.} {\ninerm
We give formal content to some concepts, naturally stemming from consistent history approach (CHA), which are 
not formalized in the standard formulation of the theory.
The outcoming (extended) conceptual basis is used to perform a
formal, conceptually transparent analysis of some debated questions in CHA.
As results, the problems raised by contrary inferences of Kent are ruled out,
whereas some prescriptions of the theory cannot be mantained.}
\vskip6mm\noindent
{\bf R}{\ninebf ESUM\'E.} {\ninerm
Nous fournissons
contenu formel \`a quelques concepts, sortant naturellement de la th\'eorie
des histoires consistantes (CHA), qui ne sont pas formaliz\'es
dans la formulation usuel. La base conceptuelle (extendue) qui en sort est us\'ee pour effectuer une
analyse formel conceptuelment transparente de quelques probl\`emes d\'ebattus dans  CHA. 
Nous avons comme resultat que le probl\`emes soulev\'es par les inferences contraires
de Kent sont r\'esolus; toutefois, des prescriptions de la th\'eorie ne peuvent pas \'etre maintenues.}
\vskip1.2cm \noindent
{\bf 1.\quad Introduction}
\vskip4.2mm\noindent
In recent works [1][2] Bassi and Ghirardi have argued that the {\sl consistent history
approach} (CHA) to quantum physics, first proposed by Griffiths [3], leads to logical 
contradiction with some assumptions ``necessary for a sound interpretation of the theory'' [1].
These criticisms to CHA generalize those moved by Kent [4], who proved that CHA ``allows contrary 
inferences from the same empirical data''. In his replies Griffiths argues that the criticisms
are the result of a misunderstanding of his theory [5].
However, Griffiths' arguments did not convince 
Bassi and Ghirardi [6]. On the other hand, also Kent [7] considered the answer of 
Griffiths and Hartle [8] to his own criticism unsatisfactory [4].
Hence, the scientific debate seems to have ended without a definitive clarification of the subject.
\par
We extend the formal apparatus of the theory, by introducing the notion of {\sl support} of a family of histories. This theoretical extension allows us to perform
a conceptually transparent analysis of the above mentioned criticisms.
The conclusions of our analysis partly agree with Griffiths; for instance the conceptual difficulties raised by contrary inferences are ruled out. But they also entail that some prescriptions of the theory cannot be mantained, in partial agreement with Bassi and Ghirardi.
\par
In section 3 we 
present a brief synthesis of the debate. To do this, we have to recall, in section 2, some
standard concepts of CHA.
In section 4 we introduce the formal tools we need for the conceptual analysis performed in sections 5 and 6.
In particular, in section 5 we deal with the criticism of Bassi and Ghirardi; while our analysis
shares Griffiths' conclusion that {\sl there is no universal truth functional in quantum history theory}, 
it entails that we cannot mantain the assumption that
{\sl every family of histories can be chosen}, in agreement with Bassi and Ghirardi.
In section 6 we show that our analysis of contrary inferences removes conceptual difficulties, also
without the necessity of adopting a weaker interpretation of perpendicular (i.e. mutually exclusive) events, 
as done by Griffiths.
\vskip4.6mm\noindent
{\bf 2.\quad Consistent History Quantum Theory}
\vskip4.2mm\noindent
Let $\cal H$ be the Hilbert space of the standard quantum description, in Heisenberg's picture,
of the physical system. Throughout this paper we assume that $\cal H$ has a finite dimension $N$.
A decomposition of the identity is a finite set
${\bf E}= \{E^{(1)},E^{(2)}, ... , E^{(k)} \}$ of projection operators such that
$E^{(i)}\perp E^{(j)}$ if $i\neq j$ and $\sum_{i=1}^{k}E^{(i)}={\bf 1}$.
Given a finite, ordered sequence of times $(t_1,t_2,..,t_n)$, for each time $t_k$
we consider a decomposition of the identity ${\bf E}_k$ whose projections represent (in Heisenberg picture) 
events which may occur at time $t_k$.
A sequence $h=(E_1,E_2,...,E_n)$ such that $E_k\in{\bf E}_k$, i.e. in 
Cartesian product ${\cal E}={\bf E}_1\times{\bf E}_2\times\cdots
\times{\bf E}_n$, is called {\sl elementary} history.
The {\sl family} $\cal C$ of {\sl histories}, generated by ${\bf E}_1,{\bf E}_2, ...,{\bf E}_n$,
is the set of all sequences $h=(E_1,E_2,...,E_n)$ such that $E_k=\sum_{\hbox{\sevenrm some
}i}E_k^{(i)}$.
Given every history $h$, we define the bounded operator $C_h=E_n\cdot E_{n-1}\cdots E_1$.
Each history
$h=(E_1,E_2,...,E_n)$ can be identified with subset ${\bf h}\subseteq\cal E$, where 
${\bf h}=\{\hat h=(\hat E_1,\hat E_2,...,\hat E_n)\in{\cal E}\mid \hat E_k\leq E_k,\quad\forall k\}$.
\par
The physical interpretation of CHA, which makes it a physical theory, is based on 
the principle which establishes that if a given family $\cal C$ 
satisfies a criterion of {\sl consistency}, then 
\item{(I)} 
{\sl the set of all elementary histories of 
$\cal C$ is a ``sample space of mutually exclusive elementary 
events, one and only one of which occurs''
[8].}
\item{}
{\sl
The occurrence of history $h$ means that an elementary history $\hat h\in{\bf h}$ occurs.
}
\vskip4.2mm\noindent
The basic physical notion of the theory is that of {\sl occurrence of a history}, 
whose meaning is the following.
\item{(O)}{
A given history  $h=(E_1,E_2,...,E_n)$
{\sl occurs} if all events $E_1,E_2,...,E_n$
{\sl objectively} occur at respective times $t_1$, $t_2$, ..., $t_n$. 
The occurrence of a history is an objective physical fact, 
independent of the performance of a measurement which reveals this occurrence.} 
\par\noindent
Hence, in this theory measurements reveal properties already objectively possessed by the physical system.
This is remarkably different from standard quantum theory, 
which, on the contrary, can make statistical predictionts only about outcomes of measurements, and  
cannot describe the occurrence of a history $h=(E_1,E_2,...,E_n)$ 
(consider the case $[E_k,E_j]\neq{\bf 0}$).
\vskip4.2mm\noindent
The criterion for establishing whether $\cal C$ is consistent is given by the following rule of CHA.
\vskip4.2mm\noindent
R{\ninerm ULE} 1.
{\sl
A family $\cal C$ is consistent if and only if it is weakly decohering, i.e., if
$Re(Tr(C_{h_1}C_{h_2}^\ast))=0$ for all $h_1,h_2\in\cal E$, $h_1\neq h_2$.}
\vskip4.2mm\noindent
The occurences of histories are the empirical facts the theory is concerned with; 
CHA postulates that their statistics obeys the following rule.
\vskip4.2mm\noindent
R{\ninerm ULE} 2.
{\sl
If $\cal C$ is consistent, then number $p(h)={1\over N}Tr(C_hC_h^\ast)$ is the probability 
of occurrence of history $h$
for every $h\in\cal C$.}
\vskip4.2mm\noindent
It can be proved that if $\cal C$ is weakly decohering then, coherently with statement (I),
for every $h\in\cal C$ we have $p(h)=\sum_{\hat h\in{\bf h}}p(\hat h)$. 
\vskip4mm
Following [9], 
to make inferences about occurrences of histories,
the {\sl initial data}, i.e. the available information about the physical system, 
must be expressed as sentences involving histories of a {\sl consistent family} 
$\cal C$, and assumed as {\sl true}, i.e. objectively realized sentences. 
At this point it is possible to derive conclusions by making logical reasonings in which histories of 
$\cal C$ are regarded, according to (I), as {\sl events of a classical sample space}. According to CHA these 
conclusions have to be regarded as empirically valid sentences.
However, another set of conclusions could be derived by using another consistent family 
${\cal C}'$ in which
the same initial data can be expressed. A conclusion of this new set could
conflict with the conclusions drawn from $\cal C$. 
In Griffiths' theory these kinds of conflicts are avoided
by means of the introduction of the following.
\vskip4mm\noindent
R{\ninerm ULE} 3.
{\sl All valid physical inferences are those obtained by using rule 2 within a {\sl single}
consistent family $\cal C$.
In general, different conclusions drawn by using distinct consistent families do not hold together.}
\vskip4.6mm
\noindent
{\bf 3.\quad Debated questions}
\vskip4.2mm\noindent
{\sl Single family} rule 3 is at the root of the criticisms mentioned in section 1.
To describe the criticisms we make use of the contrary inferences discovered by Kent [4].
\vskip4.2mm\noindent
{\bf 3.1.\quad Contrary inferences}
\par\noindent
Kent was able to find two histories 
$h_1=(E_0,E_1,E_2)$ and $h_2=(E_0,F_1,E_2)$, with $E_1\perp F_1$, belonging to two different consistent families 
${\cal C}_1$ and ${\cal C}_2$ respectively, such that
according to the rules of CHA
$$
\cases{p(E_1\mid E_0, E_2)=1& in family ${\cal C}_1$\cr
p(F_1\mid E_0,E_2)=1& in family ${\cal C}_2$\cr
p(E_0,E_2)\neq 0& (in both families).\cr}\leqno(1)
$$
If we take as initial data the occurrence of $E_0$ and $E_2$, this becomes
a very striking clear example of the above mentioned  conflict. Indeed, these initial data are compatible 
with both families ${\cal C}_1$ and ${\cal C}_2$ (third equation in (1)).
Therefore, when history $h_0=(E_0,E_2)$ occurs we conclude that
\item{$\alpha$)}
event $E_1$ occurs at time $t_1$, by reasoning with ${\cal C}_1$, 
\item{or}
\item{$\beta$)}
event $F_1$ occurs at time $t_1$,
by reasoning with ${\cal C}_2$.
\vskip4.2mm\noindent
Since $E_1\perp F_1$ means that $E_1$ and $F_1$ are {\sl mutually exclusive} events,
according to Kent the simultaneous validity of ($\alpha$) and ($\beta$) makes it ``hard to take it 
[CHA] seriously as a fundamental theory in its present form'' [4]. Even though single family rule 3 
formally prevents contrary inferences from yielding theoretical contradiction [8], ``{\sl Any} formalism
[...] can be made free from contradiction by such a restriction'' [7]. 
\par
The answer of Griffiths ([10], Appendix A) was that the problem arises because Kent assigns
to `contrary' histories $h_1$ and $h_2$ the classical-logic meaning of word {\sl contrary}, i.e. that the 
occurrence of $h_1$ always implies the non-occurrence of $h_2$, as stated by axiom 3 in section 6 of present paper.
But, according to Griffiths, this cannot be done because $h_1$ and $h_2$ cannot be compared without 
violating single family rule 3.
\vskip4.2mm\noindent
{\bf 3.2.\quad Ordered consistency}
\par\noindent
A proposal to solve the problem is due to Kent himself. He proposed to replace the original criterion of consistency, i.e. by a more restricitive one
he called {\sl ordered consistency} [11]. Kent defined the ordering
$
h_1\leq h_2$ iff
$E_k\leq F_k$ for all $k$, where $h_1=(E_1,E_2,...,E_k,...)$ and
$h_2=(F_1,F_2,...,F_k,...)$. History $h_1$ is said to be {\sl ordered consistent} if $h_1$ belongs to a consistent family and
if $h_1\leq h_2$ implies
$Tr(C_{h_1}\rho C_{h_1}^\ast)
\leq
Tr(C_{h_2}\rho C_{h_2}^\ast)$, for every $h_2$ belonging to a consistent family.
When all histories of a consistent family $\cal C$ are
ordered consistent, then $\cal C$ is said to be {\sl ordered consistent}.
Then Kent proved that contrary inferences are forbidden if the new criterion of consistency is adopted. However, it must be noticed that the sense of the proposal of Kent was not to suggest that the ordered consistent formalism is the ``right'' interpretation of quantum theory:
``The aim here is thus not to propose the ordered consistent approach as a plausible fundamental interpretation of quantum theory, but to suggest that the range of natural and useful mathematical definitions of types of quantum history is wider than previously understood.'' [11]. In the present work a different approach to the problem of contrary inferences is followed; in section 6 we show that contrary inferences do not yield contradiction if the conceptual basis 
introduced in section 4 is adopted.
\vskip4.2mm\noindent
{\bf 3.3.\quad More general criticism}
\vskip4mm\noindent
The more general criticism raised by Bassi and Ghirardi also gives a formal content to that of Kent.
They take into account four assumptions, labelled as (a), (b), (c) and (d) in [1].
The first two, (a) and (b), essentially reflect the content of our rule 1 and rule 2.
The third assumption, (c), reflects the meaning of the notion of occurrence
of history as expressed by (O): 
\item{(c)}
The occurrence of a given history ``cannot depend from the decoherent [i.e. consistent]
family one is considering.''
\par\noindent
Bassi and Ghirardi proved that these assumptions lead to logical contradiction with the following fourth
assumption.
\item{(d)}
``Any decoherent family must be taken into account'', because ``some supporters [of CHA] insist in claiming
that there are no priviledged families.'' [1].
\vskip4.2mm\noindent
In his reply [5][12] Griffiths uses essentially two arguments.
\item{1.}
The derivation of the contradiction violates the single family rule.
\item{2.}
``The conceptual difficulty goes away if one supposes that the two incompatible frameworks 
are being used to describe two distinct physical systems that are described by the same initial data'' [13].
\vskip4.2mm
All replies have not convinced [7][6] the critics of CHA.
We shall attempt to explain synthetically the reason for such a disagreement by using example 1.
Suppose that the known data about a single physical system $s$ are that $E_0$ and $E_2$
occur.
In order to establish whether $E_1$ occurs or not, a physicist can use family ${\cal C}_1$ and,
in accordance with (1), he finds that $E_1$ occurs.
But {\sl another physicist, for the same individual system $s$}, could choose ${\cal C}_2$, 
and he must conclude that $F_1$ occurs. The fact that $E_1$ occurs or $F_1$ occurs 
seems to depend on the physicist's choice of family ${\cal C}_1$ or ${\cal C}_2$. 
But, according to CHA itself, the occurrence of a 
history, once established by means of the theory, is an {\sl objective} fact.
As a consequence, both $E_1$ and $F_1$ should occur. 
But this final conclusion is rejected by everybody
because $E_1\perp F_1$.
\par
We see that replies 1 and 2 above do not provide a satisfactory answer to the problem.
Therefore, the question 
\item{\qquad}{\sl
what is the event which occurs for this $s$,
$E_1$ or $F_1$?}
\par\noindent
remains open.
\vfill\eject
\noindent
{\bf 4.\quad Conceptual basis}
\vskip4.2mm\noindent
To begin our analysis, we take into account basic principles (I) and (O).
Principle (I) entails that, given a consistent family $\cal C$,
each time the fact that `an elementary history of $\cal C$ occurs and all other
elementary histories do not occur' objectively takes place, this holds for one individual concrete
sample of the physical system.   
Then, for every family $\cal C$ we can postulate the existence of a set
$c({\cal C})$, whose elements represent all concrete physical systems,
such that for each individual $s\in c({\cal C})$ 
every history of $\cal C$ either occurs or does not occur.
In CHA language, an individual concrete system $s$ belongs to $c({\cal C})$ 
if and only if every history $h\in\cal C$ either occurs or does not occur ({\sl makes sense}) for this $s$.
The introduction of set $c({\cal C})$, we shall call {\sl support of} $\cal C$, allows us to formally express the consistency of a family 
by means of a simple definition.
\vskip4.2mm\noindent
D{\ninerm EFINITION} 1.
{\sl A family $\cal C$ is consistent if and only if $c({\cal C})\neq\emptyset$.}
\vskip4.2mm\noindent
Now we shall establish, in a coherent fashion, basic principle
(P), and general Axioms 1 and 2.
Let $s$ be an individual physical system. If $s\notin c({\cal C})$,
then a history of $\cal C$ does not necessarily make sense for $s$, 
even if $\cal C$ is consistent.
In this case to ask for the occurrence of a history $h\in\cal C$ is generally
speaking as meaningless, 
as, for instance, to ask for the political tendency of an electron.
Then, the following principle must hold:
\vskip4.2mm\noindent
\item{(P)}{\sl
the (sufficient) condition which makes the conclusions of logical reasonings
based on a family $\cal C$ valid is
$$
s\in c({\cal C}).\eqno(2)
$$}
\par\noindent
Now we introduce the first formal axiom.
\vskip4.2mm\noindent
A{\ninerm XIOM} 1.
{\sl Let ${\cal C}_1$, ${\cal C}_2$
be two families of histories. Then
$$
{\cal C}_1\subseteq {\cal C}_2\quad\hbox{implies}\quad c({\cal C}_2)\subseteq c({\cal C}_1).
$$}
Let us explain why Axiom 1 should hold.
Since  the validity of Axiom 1 is obvious
when $c({\cal C}_2)=\emptyset$, we consider the case in which $c({\cal C}_2)\neq\emptyset$.
If $s\in c({\cal C}_2)$, then there is an elementary history $h_2$ of ${\cal C}_2$ which occurs,
and all other elementary histories of ${\cal C}_2$ do not occur for this $s$. 
From ${\cal C}_1\subseteq{\cal C}_2$ it follows that all elementary histories of ${\cal C}_1$ 
form a set, denoted by ${\cal E}_1$, of albeit non-elementary histories of ${\cal C}_2$
(${\cal C}_1$ is a coarser graining than ${\cal C}_2$).
Only one history $h_1$ among those of ${\cal E}_1$ occurs in correspondence with this $s$, because there is a unique 
$h_1\in{\cal E}_1$ such that $h_2\in{\bf h}_1$, and all other $h\in{\cal E}_1$ do not occur (see (I)).
Therefore, it is possible to state that only one elementary history of ${\cal C}_1$ occurs
and all the others do not occur for this individual system $s$, thus $s\in c({\cal C}_1)$.
\vskip4mm\noindent
Now we proceed to state axiom 2.
If $h\in\cal C$, by $c_1(h;{\cal C})$ (resp., $c_0(h;{\cal C})$) we denote
the subset of $c({\cal C})$ whose elements are the systems for which $h$ occurs (resp., does not occur).
It is obvious that
$$
c_0(h;{\cal C})=c({\cal C})\setminus c_1(h;{\cal C})\quad\hbox{and}\quad
c({\cal C})=\cup_{h\in\cal E}\; c_1(h;{\cal C}).
\eqno(3)
$$
We adopt the notion of occurrence of history expressed by (O), according to which the occurrence 
of $h=(E_1,E_2,...,E_n)$ is equivalent to the occurrences of all events $E_1$, $E_2$, ...$E_n$, 
without making reference to the family which $h$ belongs to. Coherently, we state the further following axiom.
\vskip4mm\noindent
A{\ninerm XIOM} 2.
{\sl If $h\in{\cal C}\cap{\cal C}'$, then 
$c_1(h;{\cal C})\cap c_0(h;{\cal C}')=c_1(h;{\cal C}')\cap c_0(h;{\cal C})=\emptyset$.}
\vskip4mm\noindent
In other words, $h$ cannot both occur as history of $\cal C$ and do not occur 
as history of ${\cal C}'$, for the same system $s$.
\par
Now we use axioms 1 and 2 to introduce the notion of {\sl truth functional}
stemming from our approach.
Let $X$ be a set of histories. The {\sl family generated by $X$} is
${\cal C}(X)=\cap_{X\subseteq {\cal C}}\;{\cal C}$. For instance, family ${\cal C}(\{h\})$
generated by a single
history $h=(E_1,E_2,...,E_n)$ is the
family having ${\cal E}(h)=\{(F_1,F_2,...,F_n)\mid F_k\in\{E_k,E_k'\}\}$
as set of elementary histories; indeed, 
$h\in\cal C$ implies ${\cal C}(\{h\})\subseteq\cal C$.
\par
Through Axiom 1 we find that $h\in\cal C$ implies 
$c({\cal C})\subseteq c({\cal C}(\{h\}))=\cup_{h\in{\cal C}}\; c({\cal C})$. 
Therefore, $c(h)\equiv c({\cal C}(\{h\}))$ is the set of all concrete physical 
systems for which single history $h$ occurs or does not occur ($h$ makes sense).
By $c_1(h)$ (resp., $c_0(h)$)we denote the subset of those systems for 
which $h$ occurs (resp., does not occur). Of course
$$
c_0(h)=c(h)\setminus c_1(h),\qquad c_1(h)=c(h)\setminus c_0(h).
\eqno(4)
$$
Then for each physical system $s$, we can define the mapping 
$$t_s:\cup_{s\in c({\cal C})}\;{\cal C}\;\to\;\{ 0,1\},\quad t_s(h)=\cases{1&if $s\in c_1(h)$\cr
0&if $s\in c_0(h)$,\cr}\eqno(5)
$$
called {\sl truth functional} relative to $s$.
If $t_s(h)=1$, then $h$ occurs as history of ${\cal C}(\{h\})$, and hence it occurs as history of whatever
family $\cal C$ such that $h\in\cal C$ and $s\in c({\cal C})$. 
\par  
In order to perform our conceptual analysis of CHA, we have to consider the formal tools just established, together
with standard axioms of CHA, which we here formulate as 
axioms CHA.1 and CHA.2.
\par\noindent
A{\ninerm XIOM} CHA.1. {\sl A family $\cal C$ is consistent if and only if it is weakly 
decohering}.
\par\noindent
A{\ninerm XIOM} CHA.2. {\sl Let $\cal C$ be a consistent family. If
$h\in\cal C$, then $p(h)={1\over N}Tr(C_hC_h^\ast)$ is the probability
of occurrence of history $h$.}
\vskip4mm
The existence of (non-empty) set $c({\cal C})$
for every (consistent) family $\cal C$ is a logical consequence of the notion (O) of occurrence of history.
Whether a given concrete sample $s$ of the physical system belongs to $c({\cal C})$ or not is a question the physicist can try to answer by using the tools provided by the theoretical apparatus, as the axioms, and the initial data at his disposal;
for instance, see our discussion of the problem of contrary inferences in section 6. However, due to the intrinsically 
stochastic character of CHA [13], a general characterization  of $c({\sl C})$ cannot be given, though its existence cannot 
be denied without affecting th {\sl ojectivity of the occurrences of histories}, which is one of
the peculiar basic ideas of CHA. 
\par
On the other hand, we encounter a similar situation in other valuable physical theories. 
For instance, classical statistical mechanics assumes
that many micro-states correspond to one macro-state, and that the micro-state of an individual system (e.g. a gas) in a known macro-state is unique at a particular time $t$. While such an assumption is very useful in developing the theory, in general the theory itself is not able to establish the particular micro-state the system occupies when its macro-state is known.
\vskip4.2mm\noindent
{\bf 5. \quad Conceptual analysis}
\vskip4.2mm\noindent
Now we shall demonstrate explicitly that our analysis leads to the conclusion that
assumption (d) in section 3 does not hold.
Assumption (c) still holds, and we get a more precise understanding of it. 
In so doing, we can also point out how the standard interpretation should be modified.
\par
First, we consider assumption (c). Because of Axiom 2 and (5) it is clear that we have to
agree with the following idea 
expressed by Bassi and Ghirardi: ``Does the truth value of the considered 
history depend on the decoherent family to which it may belong? We think that the answer
must be ``no''. '' [6].
However, now we have a deeper understanding. The fact that a given history $h_0$ occurs 
(or does not occur) for a physical system $s_0$ entails that a family ${\cal C}_0$ exists such that 
$$
h_0\in{\cal C}_0\quad\hbox{and}\quad s_0\in c({\cal C}_0).\eqno(6)
$$
For instance, in virtue of Axiom 1, family ${\cal C}_0={\cal C}(\{h_0\})$ fullfils these requirements. 
All histories of ${\cal C}_0$ make sense for $s_0$. However, 
the eventuality that for a given system $s$ history $h\in{\cal C}$ occurs, where ${\cal C}$ is a consistent family,
but $s\notin c ({\cal C})$ is logically possible; remark 3 in section 6 provides an explicit example. 
In general, therefore, the fact that a family ${\cal C}$ is consistent, and $h\in{\cal C}$
does not imply that the inferences obtained by means of reasonings based on ${\cal C}$ hold
for an arbitrary physical system $s$ for which $h$ occurs.
\par
Furthermore, assumption (c) cannot be generally interpreted in the sense that
truth functional $t_s$ must be defined on all consistent families. 
Indeed, set ${\cal D}_s$ which $t_s$ is defined on, is just ${\cal D}_s=\cup_{s\in c({\cal C})}\cal C$.
Therefore, we also agree with Griffiths' conclusion [12] 
stating that in quantum physics {\sl there is no universal truth functional}.
\vskip4.2mm
Now we come to assumption (d). Once again, principle (P) denies that all consistent families
are valid bases for quantum reasonings. Only if a family $\cal C$ satisfies (2), can the inferences based on $\cal C$ 
be considered valid.
However, the failure of assumption (d) does not yield conceptual difficulties, because  
whether a family can be used or not depends on condition (2), rather than on the physisist's (subjective) choice.
Eventhough a general criterion for establishing whether (2)holds is not available, under certain circumstances
we can state that (2) certainly holds. For instance, suppose that the following statement holds for $s$: 
``history $h_0$ occurs'' ({\sl initial data}). Then $s\in c_1(h_0)$, and hence 
$s\in c({\cal C}(\{h_0\}))$, are true statements. Thus, all sentences obtained by 
logical deductions based on ${\cal C}(\{h_0\})$ hold for $s$.
\par
Family ${\cal C}(\{h_0\})$ admits several refinements, i.e. various families $\cal C$ may exist such that
${\cal C}(\{h_0\})\subseteq\cal C$. Deductions based on a refinement $\cal C$ 
in general do not hold for $s\in c({\cal C}(\{h_0\}))$; indeed, as already argued,
${\cal C}(\{h_0\})\subseteq\cal C$ implies, by Axiom 1, 
$c({\cal C})\subseteq c({\cal C}(\{h_0\}))$, and therefore $s\in c(\cal C)$ does not necessarily follow.
Of course, all inferences obtained by reasoning with ${\cal C}(\{h_0\})$ can be obtained
with $\cal C$, since ${\cal C}(\{h_0\})\subseteq\cal C$, therefore the two set of sentences cannot contradict
each other, but inferences involving histories in
${\cal C}\setminus{\cal C}(\{h_0\})$ in general do not make sense if $s\in c({\cal C}(\{h_0\}))\setminus c({\cal C})$.
The whole argument immediately extends to the more general case in which the data consist in the occurrence of a given set $X$ of histories, by replacing ${\cal C}(\{h\})$ with ${\cal C}(X)$.
\par
Thus, there is a profound difference between the ``coarsest'' family ${\cal C}(X)$ compatible with the available 
initial data 
relative to a physical system $s$ and any refinement $\cal C$ of ${\cal C}(X)$:
conclusions drawn by using ${\cal C}(X)$ are {\sl true}, i.e. {\sl objective facts},
whereas the truth of conclusions obtained from a refinement is not ensured
by the truth of the data ($X$) alone. 
For instance, if a reasoning made in ${\cal C}(X)$ leads us to infer that a certain history 
$h_1\in{\cal C}(X)$ occurs, such an occurrence must be considered an objective fact.
On the contrary, when the occurrence of a history $h_2\in{\cal C}\setminus{\cal C}(X)$
is inferred by a reasoning made in $\cal C$, such an occurrence cannot be considered certain.
\vskip4.2mm\noindent
R{\ninerm EMARK} 1.
Griffiths puts forward a precise strategy for choosing the family to be used:
``Use the smallest, or coarsest framework which contains both the initial data and the
{\sl additional properties of interest in order to analyse the problem}.'' [13].
In the present approach Griffiths' strategy is not always valid. Indeed, coarsest family $\cal C$ containing,
besides the initial data, the properties of interest as well,  does not satisfy (2) in general.
\vskip4mm
Thus, a conclusion of our analysis is that assumption (d) cannot be mantained, 
while assumption (c) holds.
\vskip4mm\noindent
R{\ninerm EMARK} 2. 
Kent considered single family rule 3 an unnatural expedient to avoid contradiction [7] and this was at root
of the criticisms;
we see that that it is sufficient our quite natural 
principle (P) to avoid conflicts between
conclusions drawn from different families. However,
since in Griffiths' theory the conclusions drawn in different families 
hold together in the case that these families are
`compatible', we recall the definition of compatibilty.
\vskip4mm\noindent
D{\ninerm EFINITION} 2. { \sl Two consistent families ${\cal C}_1$ and ${\cal C}_2$ are compatible if
a third consistent family $\cal C$ exists such that ${\cal C}_1\cup {\cal C}_2\subseteq\cal C$.}
\vskip4mm\noindent
Compatibility implies $c({\cal C})\subseteq c({\cal C}_1)\cap c({\cal C}_2)$; therefore, according 
to our approach, the conclusions
drawn from ${\cal C}_1$ and ${\cal C}_2$ hold together only for those systems $s$ such that
$s\in c({\cal C}_1)\cap c({\cal C}_2)$, in particular if $s\in c({\cal C})$;
on the contrary, if $s\in c({\cal C}_1)$ and $s\notin c({\cal C}_2)$,
then a conclusion drawn from ${\cal C}_2$ does not necessarily hold for this $s$
(see remark 3 in the next section for a concrete example). 
\vskip4.6mm
\noindent
{\bf 6. \quad Re-interpreting contrary inferences}
\vskip4.2mm\noindent
It is worthwhile seeing whether the particular situation of contrary inferences, 
described in example 1, can be interpreted
without encountering conceptual difficulties.
\vskip4.2mm\noindent 
Let $E$ and $F$ be two mutually orthogonal projections. 
The family generated by $E$ and $F$, i.e. ${\cal C}(\{E,F\})$, has 3 elementary (one-event) histories: ${\cal E}=\{ E,F,G={\bf 1}-(E+F)\}$; it is the smallest family containing $E$ and $F$. Then, whenever both $E$ and $F$ make sense, all 
histories in ${\cal C}(\{E,F\})$ must make sense too.
Therefore, we can state the following proposition.
\vskip4.2mm\noindent
P{\ninerm ROPOSITION} 1.
{\sl If $E\perp F$, then
$s\in c(E)\cap c(F)$ implies $s\in c({\cal C}(\{E,F\}))$ and, therefore,
$c_1(E)\cap c_1(F)=\emptyset$.}
\par\noindent
Proposition 1 says that two perpendicular projections represent mutually exclusive events.
\vskip4mm\noindent
Kent's example 1 would be a proof of a contradiction 
if ${\cal C}_1$ and 
${\cal C}_2$ were {\sl compatible} families. Indeed, 
if ${\cal C}_1$ and ${\cal C}_2$ were compatible, there would be a consistent family $\cal C$
such that ${\cal C}_1\cup {\cal C}_2\subseteq\cal C$. Then, from axiom 1
$$
c({\cal C})\subseteq c({\cal C}_1)\cap c({\cal C}_2)\eqno(7)
$$
would follow; furthermore $h_0=(E_0,E_2)\in\cal C$.
By $p(E_0,E_2)\neq 0$ in (1) we find that there exists $\hat s\in c_1(h_0)\cap c({\cal C})$.
By (7), $\hat s\in c({\cal C}_1)\cap c({\cal C}_2)$. Therefore, by (1) we should conclude
$\hat s\in c_1(E_1)$ and $\hat s\in c_1(F_1)$.
Thus we have a contradiction with $c_1(E_1)\cap c_1(F_1)=\emptyset$ following from proposition 1, since $E_1\perp F_1$.
\vskip4mm
However, ${\cal C}_1$ and ${\cal C}_2$ in example 1 are not compatible families, 
therefore this argument does not apply. 
On the contrary,
the occurrence of history $h_0=(E_0,E_2)$ does not give rise to conceptual difficulties.
The condition to be satisfied to avoid the contradiction is
$$
c_1(h_0)\cap c({\cal C}_1)\cap c({\cal C}_2)=\emptyset.
\eqno(8)
$$
Indeed, if $s\in c_1(h_0)\cap c({\cal C}_1)\cap c({\cal C}_2)\neq\emptyset$,
then $s\in c_1(E_1)\cap c_1(F_1)\neq \emptyset$ would follow from (1), in contradiction with Prop.1.
\par
If ${\cal C}_1$ and ${\cal C}_2$ are not compatible, (8) is logically consistent 
with the occurrence of $h_0$.
Indeed, from
$$
h_0\in{\cal C}_1\cap{\cal C}_2, \eqno(9)
$$
by axiom 1 we get
$$
c({\cal C}_1)\cup c({\cal C}_2)\subseteq c(h_0),
$$
which is consistent with (8).
In particular,
when $h_0$ occurs, i.e. $s\in c_1(h_0)\subseteq c(h_0)$, we have 3 distinct possibilities, all
consistent with (8); namely
\item{p)}
$s\in c({\cal C}_1)$, and then $E_1$ occurs. In such a case consistency with (8) requires that
$s\notin c({\cal C}_2)$, therefore the second equation in (1), stating that
`$h_0$ occurs implies $F_1$ occurs', does not hold for this $s$. 
There are two possibilities regarding the occurrence of $F_1$:
\item{}
p$_1$) $s\in c(F_1)\setminus c({\cal C}_2)$. This possibility is consistent because
$F_1\in{\cal C}_2$ implies $c({\cal C}_2)\subseteq c(F_1)$. By axioms 2 and proposition 1
we must conclude that $F_1$ does not occur, i.e. $s\in c_0(F_1)$;
\item{} p$_2$)
$s\notin c(F_1)$, therefore $F_1$ does not make sense, i.e. it neither occurs nor does not occur.
\item{q)}
$s\in c({\cal C}_2)$, and then $F_1$ occurs. In such a case 
we have for $E_1$ the same conclusions of item (a) for $F_1$.
\item{r)}
$s\notin c({\cal C}_1)\cup c({\cal C}_2)$. In this case no inference about $E_1$ or $F_1$ 
can be drawn from the data and (1).
\vskip4.2mm\noindent
Which of the alternative, and mutually exclusive, possibilities (p), (q) and (r)
above is actually realized with our initial data ($s\in c_1(h_0)$) is a question which cannot 
be answered without further data. 
\vskip4.2mm\noindent
R{\ninerm EMARK} 3.
The foregoing argument provides a concrete example in which a statement drawn from a
consistent family (${\cal C}(\{h_0\})$) holds, whereas another statement drawn from another 
family compatible with ${\cal C}(\{h_0\})$
does not hold. Indeed, statement ``$h_0$ occurs'' must hold for some $s\in c({\cal C}(\{h_0\}))$, because 
$p(h_0)\neq 0$, whereas at least one of the statements ``$E_1$ occurs'' or ``$F_1$ occurs'', drawn in correspondence
with initial datum ``$h_0$ occurs'' from ${\cal C}_1$ or ${\cal C}_2$, must not hold, where both
${\cal C}_1$ and ${\cal C}_2$ are compatible with $c({\cal C}(\{h_0\}))$.
\vskip4mm\noindent
R{\ninerm EMARK} 4.
In [13] Griffiths presents another argument for showing the absence of any contradiction in CHA. In discussing
the 3-boxes paradox, quite equivalent to contrary inferences, he shows that the contradiction arises
when Axiom 3 below is assumed to hold. 
\vskip4mm\noindent
A{\ninerm XIOM} 3. {\sl If $E\perp F$, then
\item{i)}
$s\in c(E)\cap c(F)$ implies $s\in c({\cal C}(\{E,F\}))$,
\item{ii)}
$c_1(E)\subseteq c_0(F)$.}
\vskip4mm\noindent
This axiom relies more on implications drawn from perpendicularity, than proposition 1.
Griffiths argues that the validity of this axiom is misleading in CHA, and therefore
the contradiction disappears once Axiom 3 is ruled out.
Our consistent interpretation of contrary inferences actually does not make use of Axiom 3.
This may therefore give rise to the suspicion 
that our solution of contrary inferences works 
only because we have not assumed Axiom 3.
On the contrary, 
our argument above can be repeated along the same lines with Axiom 3 instead of proposition 1. 
The results turn out to be the same, with
the only difference that possibility (p$_2$) in item (p) can no longer occur.
\vfill\eject\noindent
{\bf REFERENCES}\vskip4.2mm\noindent
\item{[1]} A. Bassi, G.C. Ghirardi, Phys. Letters A {\bf 257}, 247 (1999) \par\noindent
\item{[2]} A. Bassi, G.C. Ghirardi, J.Stat.Physics {\bf 98}, 457 (2000)\par\noindent
\item{[3]} R.B. Griffiths, J.Stat.Physics {\bf 36}, 219 (1984)\par\noindent
\item{[4]} A. Kent, Phys.Rev.Letters {\bf 78}, 2874 (1997)\par\noindent
\item{[5]} R.B. Griffiths, J.Stat.Physics {\bf 99}, 1409 (2000)\par\noindent
\item{[6]} A. Bassi, G.C. Ghirardi, Phys.Letters A {\bf 265}, 153 (2000)\par\noindent
\item{[7]} A. Kent, Phys.Rev.Letters {\bf 81}, 1982 (1998)\par\noindent
\item{[8]} R.B. Griffiths, J.B. Hartle, Phys.Rev.Letters {\bf 81}, 1981 (1998)\par\noindent
\item{[9]} R.B. Griffiths, Phys. Rev. A {\bf 54}, 2759 (1996)
\item{[11]} A. Kent, in {\sl Relativistic quantum measurement and decoherence}, H.P. Breuer and F Petruccione (Eds.)
Lecture Notes in Physics {\bf 559}, 93 (2000)
\item{[10]} R.B. Griffiths, Phys. Rev. A {\bf 57}, 1604 (1998)
\item{[12]} R.B. Griffiths, Phys. Letters A {\bf 265}, 12 (2000)
\item{[13]} R. Griffiths, {\sl Consistent quantum theory} (Cambridge University Press, Cambridge 2002)

\bye